\documentclass[useAMS,usenatbib]{mnras}
\usepackage{mathtext,amssymb,amsmath}
\usepackage{epsfig}
\usepackage{graphics}
\usepackage{url}
\usepackage{adjustbox}

\usepackage{times}

\usepackage[T1]{fontenc}
\newcommand{\be}{\begin{equation}}
\newcommand{\ee}{\end{equation}}
\newcommand{\beq}{\begin{eqnarray}}
\newcommand{\eeq}{\end{eqnarray}}

\newcommand{\Msun}{{\rm M}_{\odot}}

\newcommand{\fc}{f_{\rm c}}

\def\ergcm2s{{\rm erg\,cm^{-2}\,s^{-1}}}

\title[Direct cooling tail method]{The direct cooling tail method for X-ray burst analysis 
to constrain neutron star masses and radii}

\author[V. F. Suleimanov et al.]{Valery F. Suleimanov,$^{1,2}$\thanks{E-mail: suleimanov@astro.uni-tuebingen.de}
Juri Poutanen,$^{3,4}$
Joonas N\"attil\"a,$^{3,4}$
Jari J.~E. Kajava,$^{3,5}$
\newauthor
\fbox{Mikhail G. Revnivtsev$^6$},
and Klaus Werner$^{1}$
\\
$^{1}$Institut f\"{u}r Astronomie und Astrophysik, Kepler Center for Astro and
Particle Physics, Universit\"{a}t T\"{u}bingen,\\ Sand 1, D-72076 T\"{u}bingen,
Germany\\
$^2$Kazan (Volga region) Federal University,  Kremlevskaya str. 18, Kazan 420008, Russia\\
$^{3}$Tuorla Observatory, Department of Physics and Astronomy, University of Turku, V\"{a}is\"{a}l\"{a}ntie 20, 
FIN-21500 Piikki\"{o}, Finland\\
$^{4}$Nordita, KTH Royal Institute of Technology and Stockholm University, Roslagstullsbacken 23, SE-10691 Stockholm, Sweden\\
$^{5}$European Space Astronomy Centre (ESA/ESAC), Science Operations
Department, 28691 Villanueva de la Ca\~{n}ada, Madrid, Spain\\
$^6$Space Research Institute, Russian Academy of Sciences, Profsoyuznaya 84/32,
117997 Moscow, Russia}

\pubyear{2016}

\begin{document}
\label{firstpage}
\pagerange{\pageref{firstpage}--\pageref{lastpage}}
\maketitle

\begin{abstract}
Determining  neutron star (NS) radii and masses  can help to understand the properties of matter  at supra-nuclear densities. 
Thermal emission during thermonuclear X-ray bursts from NSs in low-mass X-ray binaries 
provides a unique opportunity to study NS parameters, because of the high fluxes, large luminosity variations 
and the related changes in the spectral properties. 
The standard cooling tail method uses hot NS atmosphere models to convert
the observed spectral evolution during cooling stages of X-ray bursts 
to the Eddington flux $F_{\rm Edd}$ and the stellar angular size $\Omega$. 
These are then translated to the constraints on the NS mass $M$ and radius $R$.
Here we present the improved, direct cooling tail method that generalises the standard approach. 
First, we adjust the cooling tail method to account for the bolometric correction to the flux. 
Then, we fit the observed dependence of the blackbody normalization on flux with a theoretical model 
directly on the  $M-R$ plane by interpolating theoretical dependences to a given gravity, hence 
ensuring only weakly informative priors for $M$ and $R$ instead of $F_{\rm Edd}$ and $\Omega$.
 The direct cooling method is demonstrated using a photospheric radius expansion burst from SAX\,J1810.8--2609,
which has happened when the system was in the hard state. 
Comparing to the standard cooling tail method, the confidence regions are shifted by 1$\sigma$  
towards larger radii,  giving $R=11.5-13.0$~km at $M=1.3-1.8\,\Msun$ for this NS. 
%Our fundings support rather {\bf } 
%equation of state of cold dense matter in agreement with the results from nuclear physics.  
\end{abstract}


%%%%%%%%%%%%%%%%%%%%%%%%%%%%%%%%%%%%%%%%%%%%%%%%%%

%%%%%%%%%%%%%%%%% BODY OF PAPER %%%%%%%%%%%%%%%%%%
\section{Introduction}

The internal structure of neutron stars (NSs) is not exactly known because of 
uncertainties in the matter equation of state in NS cores, where the matter density
is larger than nuclear density \citep*{HPY:07}. There are many theoretical models of the
equation of state, which predict various mass--radius dependences
for NSs  \citep{LP07}. Therefore, the possible dense matter equation of states could be
constrained using the measurements of the NS radii $R$ and masses $M$ \citep[see, e.g.,][]{LS:14}.
 
NS masses are accurately measured in close binary systems with radio pulsars.   
These works establish a relatively narrow range of NS masses, between 1.2 and 2 $\Msun$ \citep[see, e.g.,][]{TC99,OPN12,Kiziltan13,Anton.etal:13}. 
Within the next 20 years, radius measurement with $\sim5\%$ accuracy in such systems will be possible only for the double pulsar, PSR J0737--3039, 
via its moment of inertia  \citep{Kramer09}.  
Spectral studies of the thermal emission from NSs provide a framework for NS radii determination. 
Although the observed X-ray spectra are thermal, a simple fit with the blackbody gives too small NS radii \citep[see, e.g. ][]{PL:09,Klochkov.etal:13}.  
This is explained by the fact that the  spectrum emergent from the gaseous NS atmospheres is significantly harder than the blackbody 
with equivalent effective temperature $T_{\rm eff}$ 
and the corresponding flux level in the Rayleigh-Jeans part of the spectrum is lower \citep[see e.g.][]{Zavlin.etal:96,sw:07}. 
Therefore, to achieve the same bolometric flux and to fit the observed spectra with the atmospheric models, a 
larger NS apparent radius is required  (see e.g. \citealt{Zavlin.etal:98,HH:09}; \citealt{Klochkov.etal:13}).
The atmosphere models fits give the ratio of the apparent NS radius  at  infinity $R_\infty=R(1+z)$ 
(here $z$ is the surface redshift) to the source distance $D$. 
The known  distance to the NSs in globular clusters or in supernova remnants allows then 
to get $R_\infty$ and constrain $R$ \citep{HR06,Heinke.etal:14,Gulliot.etal:13,Klochkov.etal:13,Klochkov.etal:15}.
An important part of this method is the availability and reliability of the  NS model atmospheres and their emergent spectra. 

One of the possible ways to obtain simultaneous  $M-R$ constraints is via fast X-ray timing \citep{watts16} using 
waveforms of oscillations during thermonuclear type I X-ray bursts \citep{Lo13,ML15} or of the persistent emission of accreting millisecond pulsars  \citep{PG03}.
Alternatively,  we can use spectral information from X-ray bursting NSs in low-mass X-ray binaries 
\citep[see reviews by][]{LvPT93,Ozel13,Miller13,ML16,Suleimanov16EPJA}. 
Both the observed X-ray burst spectra \citep{gallow08,Guver.etal:12,Worpel.etal:15} and the 
theoretical model atmosphere  spectra are reasonably well fitted by a diluted blackbody   \citep{SPW11,SPW12}. 
The flux escaping the atmosphere can be represented as  $F_{E} \approx \pi w  B_{E} (T_{\rm c}= \fc T_{\rm eff})$ 
(where $T_{\rm c}$ is  the colour temperature and the dilution factor $w$ is related to the colour  correction factor 
$\fc$ as $w\approx \fc^{-4}$).  
If  $\fc$ is known, we can estimate $R_\infty$ from the measurements of the blackbody radius $R_{\rm bb}$ (for given distance) through $R_\infty=R_{\rm bb}\fc^2$. 
Of particular interest are the so called photospheric radius expansion (PRE) bursts, whose luminosity exceeds the Eddington limit $L_{\rm Edd}$. 
These bursts allow  to obtain an additional limitation on NS $M$ and $R$  
either via the ratio of the fluxes at the PRE phase and the touchdown (when the photosphere touches the NS surface)
or via estimation of the Eddington flux when the atmosphere reaches the Eddington limit   \citep{Ebisuzaki87,Damen:90,vP:90,LvPT93}. 

 \citet{Ebisuzaki87} proposed to fit the observed evolution of the colour temperature with luminosity in the cooling tail 
with his theoretical atmosphere models to find the Eddington temperature $T_{\rm Edd,\infty}$, which is  
the redshifted effective temperature corresponding to the Eddington limit at the NS surface. 
Measurements of  $T_{\rm Edd,\infty}$ gives a constraint on $M$ and $R$ \citep[see also][]{SPRW11}. 
\citet{vP:90} fitted  the  evolution of the ratio  of the observed colour temperature to $F^{1/4}$ (where $F$ 
is the observed bolometric flux) 
by theoretical models of the  evolution of  $\fc$ with luminosity (in units of the Eddington one) finding $\fc$ at 
different points of the cooling tail and using them to determine $R_\infty$. This method also allowed to check the consistency 
of $R_\infty$  measured from different points of the cooling tail. 

More recently a simplified approach was taken by 
 \citet{Ozel06} and \citet{ozel:09}. They proposed  the ``touchdown method'', where the Eddington flux is identified 
 with the touchdown flux and  the average normalization of the blackbody in the burst cooling tail was used to estimate 
 $R_\infty$ with a fixed value of $\fc$ corresponding to a low luminosity.  
For the analysis, they selected bursts which occur in the soft state of the source at high accretion rate 
and show minimum evolution of $R_{\rm bb}$  in the cooling tail. 
Such an approach was criticised for a number of reasons \citep[see][]{SLB10,SPRW11,Miller13,Poutanen.etal:14,ML16}. 
Firstly, it is not clear whether the Eddington limit  is actually reached at the touchdown. 
Secondly, the estimation of the Eddington luminosity ignores the difference between actual 
Compton scattering cross-section and the Thomson one.
Thirdly, assumption of the (nearly) constant $\fc$ as well as the selection of the bursts based 
on the constancy of observed  $R_{\rm bb}$ strongly contradicts the atmosphere models. 
Fourthly, the constraints on NS parameters are heavily dependent  on the details of the assumed distance distribution. 
Furthermore, it was shown that the touchdown method applied to the bursts happening 
in the soft state with the assumed cuts on distance, 
assumption about chemical composition and $\fc$ produce the results which have very small probability,
 i.e. the method is internally inconsistent. 
The most recent version of the touchdown method \citep{Ozel16} accounts for the deviations of the cross-section from the 
Thomson one using the approximation from \citet{SPW12}, but introduced other errors, such as considering systematic errors 
the same way as the statistical ones  \citep[see][]{ML16}. 
 
Recently we have suggested the cooling tail method for determination of NS masses and radii 
from the spectral evolution data during the cooling tail of the burst (\citealt{SPRW11}; see also Appendix A of \citealt{Poutanen.etal:14}).  
In this method, we fit the evolution of the  observed blackbody normalization with flux by the theoretical model of $\fc^{-4}$ dependence on the 
relative luminosity $\ell = L/L_{\rm Edd}$.
However, we relax the assumption often made that the Eddington limit is reached at the touchdown \citep{Ebisuzaki87,vP:90,Ozel06,Ozel16} 
and we use more data from the cooling tail to determine the Eddington flux. 
The method implicitly assumes that X-ray burst spectra as well as the model atmosphere spectra are well approximated by the  blackbody emission
and that the best-fitting blackbody flux well represents the bolometric flux of the source and the model. 

For comparison with the data, we have computed an extensive set of hot NS atmosphere models \citep{SPW11,SPW12} 
at a large grid  of $\ell$,  surface gravity $\log g$ and chemical composition. 
Models with increased abundances of heavy elements, which mimic atmospheres polluted with nuclear burning ashes, were also computed \citep{Netal15}.
Similarly to the touchdown method, the fitting procedure provides two parameters: the observed Eddington flux $F_{\rm Edd}$ (defined for Thomson opacity) 
and the ratio $R_\infty/D$. These parameters are then converted to the $M$ and  $R$ constraints assuming some probability
distribution for the distance. The cooling tail method was implemented to obtain NS mass-radius limitations for two X-ray bursting sources,
4U\,1724--307 \citep{SPRW11} and 4U\,1608--52 \citep{Poutanen.etal:14} which obtained a lower limit of $R>13$~km for 
 the mass range $1.2<M/\Msun<2$. 
However, transformation from a pair $(F_{\rm Edd},R_\infty/D)$ to $(M,R)$ is not trouble-free: because of a divergence in the Jacobian 
the resulting distribution of $M$ and  $R$ can be biased away from the ``2-Schwarzschild-radius'' line, $R=4GM/c^2$, as was shown by \citet{Ozel15}. 
This problem can be easily avoided by performing a fit to the cooling tail data directly on a grid in the $M-R$ plane  by 
interpolating the computed theoretical $\fc-\ell$ relations to the corresponding value of $\log g$. 
Such an approach within a Bayesian framework was adopted by \citet{nattila16},
allowing to  find mass-radius constraints for three bursters 4U\,1702--429, 4U\,1724--307, and SAX\,J1810.8--260.

Here we present the improved, direct cooling tail method that generalises the standard approach, 
allows deviations of $w$ from $\fc^{-4}$ and is free from the problems with the aforementioned transformation. 
In the new method, for every $(M,R)$ pair we first get the gravity $\log g$ and 
find the  relation between model dilution factor $w$ and  the colour correction factor $\fc$  on $\ell$ 
for the given  $\log g$  by interpolating these quantities on a set of pre-computed atmosphere models covering an extended 
range of  $\log g$. 
We then fit this relation to the dependence of the blackbody normalization on blackbody flux, $K - F_{\rm BB}$, 
observed in the cooling tail, using the distance as the only fitting parameter. 
The resulting  $\chi^2$ map constrains the NS  $M$ and $R$. 
We demonstrate the method using a PRE burst that occurred while SAX J1810.8--2609 was in the hard spectral state. 
 
 We note here that the reliable constraints on NS parameters can be obtained only from 
 the bursts that show evolution of the blackbody with flux consistent with the predictions of
  the atmosphere models of passively cooling NS. 
It turned out that only some bursts  happening during the low/hard state of the sources satisfy 
this requirement \citep{Poutanen.etal:14, Kajava.etal:14}. 
Interestingly, some long, very hard-state bursts show too strong spectral evolution that can 
be explained by metal enrichment of the atmosphere  \citep{SPRW11,nattila16,Kajava17}, 
but making their usage for $(M,R)$ determination difficult. 
On the other hand, none of the bursts happening in the soft, high-accretion-rate state 
shows an evolution consistent with the atmosphere model predictions \citep{Kajava.etal:14}. 
This is likely because of strong influence of the accreting matter on the atmosphere 
\citep{SPRW11,Poutanen.etal:14,Koljonen16} and/or eclipsing of the NS by the accretion disc, 
 making all  constraints coming from the soft state bursts and still using the passively cooling atmosphere
models  \citep[see e.g.][]{Ozel16}  questionable.

We also note that constraints on NS masses and radii can be obtained by direct fitting 
of the observed spectra with the atmosphere models (without going through the intermediate step of
 fitting the blackbodies to the data and the models).  
For example, \citet{Kusm:11} fitted two separate spectra of a strong X-ray burst in the ultracompact 
binary 4U\,1820$-$30 with their own helium rich model atmosphere spectra and obtained rather weak 
constraints on $M$ and $R$. They did not, however, use any information about the spectral evolution of the burst. 
On the other hand, we could  fit simultaneously all the burst spectra  getting improved $(M,R)$ constraints, 
but this is a much more demanding exercise (J. N\"attil\"a et al., in prep.).

The present paper is constructed as follows. 
In Section~\ref{s:ctm} we present the basic formulae, describe the standard cooling tail method and 
the correction to the method which accounts for the difference between the dilution factor $w$ and $\fc^{-4}$.  
In Section~\ref{sec:direct} we present the new direct cooling tail method which is implemented  
on the $M-R$ plane and does not suffer from the problems related to the divergence of the Jacobian. 
We then apply the new method to a PRE burst of SAX\,J1810.8--2609, show how the 
constraints coming from the direct cooling tail method differ from those obtained by the standard method and 
discuss possible systematic uncertainties.  
We conclude in Section~\ref{sec:summary}.

\section{Cooling tail method}
\label{s:ctm}

\subsection{Basic relations}

Let us consider a  non-rotating NS of gravitational mass  $M$ and  circumferential radius $R$, emitting 
luminosity $L$ homogeneously over its surface. 
The NS atmosphere models implicitly assume that there is no any external influence to the 
NS surface that might affect either the emission or its propagation to the observer 
(i.e. no eclipse by the accretion disc, no boundary/spreading layer). 
In addition to the chemical composition, the NS emission depends on the surface gravity
\be \label{g}
         g = \frac{GM}{R^2}(1+z) 
\ee
and the surface effective temperature given by 
\be
        \sigma_{\rm SB}\,T_{\rm eff}^4 = \frac{L}{4\pi R^2}.
\ee
Here $z$ is the  surface gravitational redshift  given by relation 
\be \label{z}
   1+z = \left(1-\frac{R_{\rm S}}{R}\right)^{-1/2} , 
\ee
$ \sigma_{\rm SB}$ is the Stefan-Boltzmann constant 
and $R_{\rm S}=2GM/c^2$ is the Schwarzschild radius. 
It is also useful to introduce the critical Eddington luminosity  and the 
corresponding effective temperature $T_{\rm Edd}$: 
\be
        L_{\rm Edd} = \frac{4\pi\,GM\,c}{\kappa_{\rm T}}(1+z) = 4\pi R^2\,\sigma_{\rm SB}T_{\rm Edd}^4 , 
\ee
where $\kappa_{\rm T}=0.2(1+X)$\,cm$^2$\,g$^{-1}$ is the Thomson electron scattering opacity and $X$ is the
hydrogen mass fraction of the atmospheric plasma.

The observed values change due to the general relativistic effects 
\be
   R_{\infty} = R\,(1+z), \quad T_{\infty} = \frac{T_{\rm eff}}{1+z}, \quad L_\infty = \frac{L}{(1+z)^2}. 
\ee
The observed Eddington luminosity and temperature are also reduced
\beq \label{eq:ledd}
         L_{\rm Edd,\infty} &= &\frac{4\pi\,GM\,c}{\kappa_{\rm T}} \frac{1}{1+z}, \\
          \label{eq:tedd}
         T_{\rm Edd,\infty} &= &\left(\frac{g\,c}{\sigma_{\rm SB}\,\kappa_{\rm T}}\right)^{1/4} \frac{1}{1+z}.
\eeq

It is well known that the majority  of the observed X-ray burst spectra   
can be fitted by the Planck function  
with two parameters, the observed colour temperature $T_{\rm BB}$ and normalization $K$ \citep{gallow08,Guver.etal:12,Worpel.etal:15}: 
\be
    F_{\rm E} \approx \pi B_{\rm E}(T_{\rm BB})\,K = \pi B_{\rm E}(T_{\rm BB})\,\frac{R_{\rm BB}^2}{D^2}.
\ee
The  observed  bolometric flux in this approximation is
\be \label{eq:Fbb}
    F_{\rm 	BB} = \sigma_{\rm SB}\,T_{\rm BB}^4\,K.
\ee
On the other hand, the observed bolometric flux is determined  as
\be  \label{eq:Finfty}
      F_\infty = \sigma_{\rm SB}\,T_\infty^4\, \Omega , 
\ee
where 
\be
\Omega  =    \frac{R_{\infty} ^2}{D^2} 
\ee
is the angular dilution factor proportional to the solid angle occupied by the NS on the sky.  
Equations (\ref{eq:Fbb}) and (\ref{eq:Finfty})  can be combined to find a 
relation between the observed effective and colour temperatures, blackbody normalization and  $\Omega$: 
\be\label{eq:TbbTinfK}
T_{\rm BB}^4\, K = T_\infty^4 \,\Omega. 
\ee

\subsection{Standard cooling tail method}

According to the NS atmosphere models \citep{SPW11,SPW12}, 
the emergent burst spectra are not real blackbodies, but they have blackbody-like shapes due to
strong interactions between electrons and photons by Compton scattering in the surface atmospheric layers. 
The model spectra can be fitted in the observed X-ray energy range by a diluted blackbody:  
\be \label{eq:fit}
     \mathcal{F}_{\rm E} \approx \pi \frac{1}{\fc^4} B_{\rm E} (T_{\rm c}), 
\ee
with the colour temperature  which is higher than the model atmosphere effective temperature  by a colour corrector factor $\fc$: 
\be
      T_{\rm c} = \fc T_{\rm eff}.
\ee 
Approximation (\ref{eq:fit}) conserves the bolometric flux
\be
   \int_0^\infty  {\fc^{-4}} \pi B_{\rm E}(T_{\rm c}) \,dE   = \fc^{-4}\sigma_{\rm SB}\,T_{\rm c}^4 = \sigma_{\rm SB}\,T_{\rm eff}^4 
\ee
and  allows us to connect the intrinsic NS effective temperature $T_{\rm eff}$ to the observed
blackbody (colour) temperature: 
\be \label{tmp}
       T_{\rm BB} = \frac{T_{\rm c}}{1+z} = \frac{\fc T_{\rm eff}}{1+z} = \fc T_\infty.
\ee 
Combining this with equation (\ref{eq:TbbTinfK}), we get 
\be \label{eq:knorm}
     K = \frac{1}{\fc^4} \Omega,  
\ee
or 
\be \label{eq:knorm-4}
K^{-1/4}= \fc A = \fc   \left(\frac{R_\infty}{D}\right)^{-1/2} . 
\ee
 
The computed $\fc$ vary  in the range 1.0--1.8, depending on the fundamental parameters
of the NS atmosphere: $T_{\rm eff}$ (or the relative NS luminosity $\ell=L/L_{\rm Edd}$), 
the surface gravity $\log g$ and the chemical composition \citep{SPW11,SPW12,Netal15}.
Equation (\ref{eq:knorm}) implies that the observed normalization $K$ has to depend on the observed bolometric flux $F_{\rm BB}$ exactly 
the same way as $\fc^{-4}$ depends on $\ell$. 
This behaviour is expected if our assumption of a passively cooling NS during the cooling tail of X-ray burst is correct. 
Indeed, X-ray bursts happening in a low/hard persistent spectral state  show the predicted $K - F_{\rm BB}$ behaviour 
\citep{SPRW11,Poutanen.etal:14, Kajava.etal:14}.
Therefore, we can fit the observed dependence $K^{-1/4} - F_{\rm BB}$ by the theoretical one $\fc - \ell$, and obtain two fitting
parameters $A$, defined through equation~(\ref{eq:knorm-4}), and the observed Eddington flux
\be
   F_{\rm Edd} = \frac{L_{\rm Edd, \infty}}{4\pi D^2}=\frac{GM\,c}{\kappa_{\rm T}\,D^2} \frac{1}{1+z}.
   \ee
They can be combined to the observed Eddington temperature, which is independent of $D$
\be \label{teddo}
  T_{\rm Edd, \infty} = AF_{\rm Edd}^{1/4} = 9.81\,A'F_{\rm Edd, -7}^{1/4}\,{\rm keV},
\ee
where $F_{\rm Edd, -7} = F_{\rm Edd} / 10^{-7}$\,erg\,s$^{-1}$\,cm$^{-2}$, and 
the normalized fitting parameter $A'=(R_\infty{\rm [km]}/D_{\rm 10})^{-1/2}$ with
$D_{\rm 10} = D/ 10$\,kpc.
The $T_{\rm Edd, \infty}$ found from observation determines a specific curve on the $M-R$ plane, 
 which allows the NS radius to be evaluated, as NS masses are in a finite range 1.2--2 $\Msun$.
This approach was successfully applied to some bursts from NSs in low-mass X-ray binaries 4U\,1724--307
\citep{SPRW11}, 4U\,1608--52  \citep{Poutanen.etal:14}, and 4U\,1702--429, 4U\,1724--307, and SAX\,J1810.8--260 \citep{nattila16}.
We note that there is a difference between the NS radius in 4U\,1724--307 obtained by \citet{SPRW11} and \citet{nattila16} because they considered different hard-state bursts. 
The burst discussed by \citet{SPRW11} has likely a metal-enriched atmosphere  \citep{nattila16}, which was not accounted for by \citet{SPRW11}.

\subsection{Importance of the spectral dilution factor}

\begin{figure} 
\begin{center}
\includegraphics[width= 1.\columnwidth]{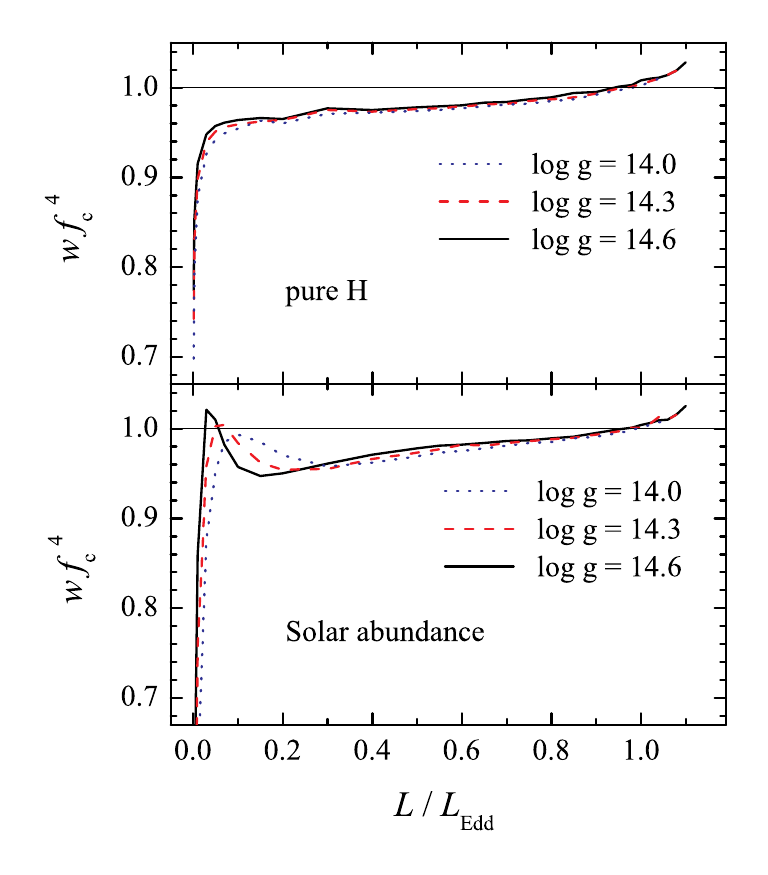}
\caption{\label{fig:bolo} 
Dependence of the correction factor $w\fc^4$ on the relative luminosity $\ell$ for various $\log g$ and two
chemical compositions.  
} 
\end{center} 
\end{figure}

The actual spectra from the NS atmosphere models have been fitted in the energy band 3--20 keV 
(corresponding to the largest effective area of the Proportional Counter Array onboard of the \textit{Rossi X-ray Timing Explorer}) 
with the two-parameter model   \citep{SPW11,SPW12,Netal15} 
\be \label{eq:fit2}
     \mathcal{F}_{\rm E} \approx w \pi B_{\rm E}(\fc T_{\rm eff}),
\ee
allowing the spectral dilution factor $w$ to be different from $\fc^{-4}$. The values $w\fc^4$ were also tabulated.
For the standard method, we used the presentation of the normalization (\ref{eq:knorm}), because  the values $w\fc^4$ are close to unity for almost  all $\ell$ 
and the deviations become significant  only at low luminosities,  $\ell \lesssim 0.1$, which are unimportant for the cooling tail method
(see Fig.~\ref{fig:bolo}).

 \begin{figure} 
\begin{center}
\includegraphics[width= 1.\columnwidth]{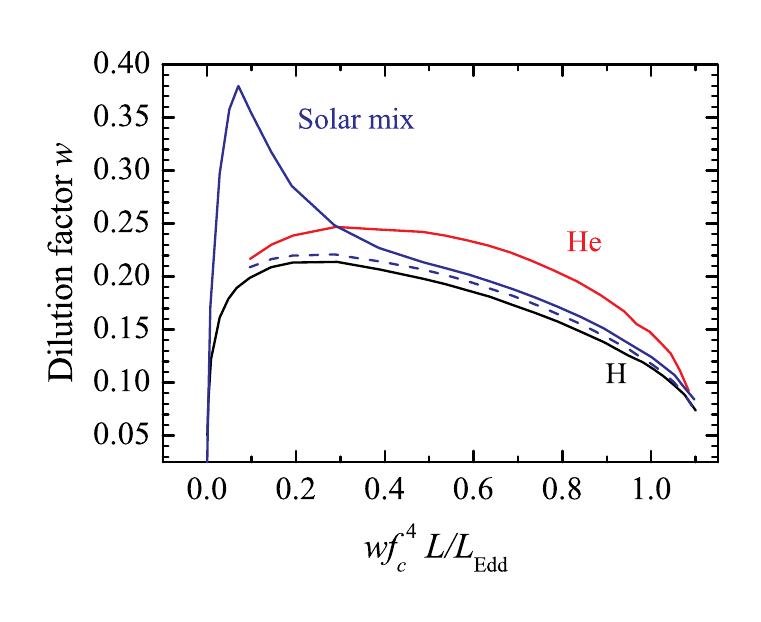}
\caption{\label{fig:dilut} 
Dependences of the dilution factor $w$ on the model  corrected relative luminosity $w\fc^4 \ell$ for fixed $\log g=14.3$ and various 
chemical compositions. The dashed curve corresponds to the solar H/He mix and reduced metal abundance ($Z=0.01Z_\odot$). 
} 
\end{center} 
\end{figure} 
 \begin{figure} 
 
\begin{center} 
\includegraphics[width= 1.\columnwidth]{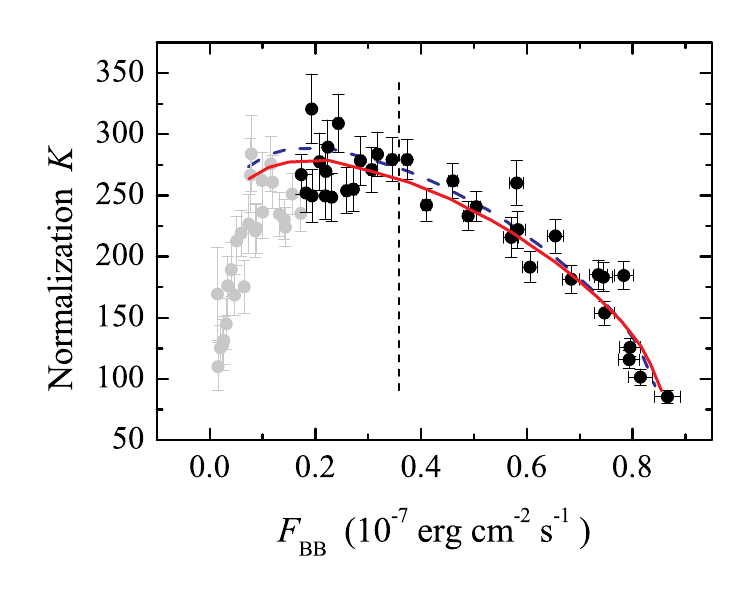}
\caption{\label{fig:ktbb} 
The observed $K - F_{\rm BB}$  dependence (circles with error bars) 
 for the cooling tail (after touchdown) of the PRE burst of SAX\,J1810.8--2609 \citep[see][]{nattila16}. 
The red solid and the blue dashed curves are 
the  best-fitting  theoretical dependences $w-w\fc^4 \ell$ for  solar H/He mix  ($Z=0.01Z_\odot$) and  $\log g = 14.3$ 
corresponding to the  data above $F_{\min}=0.2F_{\rm td}$  
(black circles) and $F_{\min}=0.4F_{\rm td}$ (points to the right of the vertical dashed line), respectively. 
 The best-fitting parameters are given in Table~\ref{table:1}.
  }
\end{center} 
\end{figure}

Let us now adjust  the cooling tail method accounting  for the correct  spectral dilution factor. 
We note that the two parameter fitting (\ref{eq:fit2}) does not conserve the surface bolometric flux
\be
     \int_0^\infty  w \pi B_{\rm E}(\fc T_{\rm eff}) \,dE = w\fc^4\,\sigma_{\rm SB}T_{\rm eff}^4,
\ee
and we need a bolometric correction factor $(w\fc^4)^{-1}$ 
 to convert the observed blackbody flux to the actual bolometric flux
\be \label{trueflux}
           F_\infty = (w\fc^4)^{-1}\, F_{\rm BB}.
\ee 
Thus, we can relate the observed black body flux to the relative luminosity as 
\be\label{eq:Fbb_l}
F_{\rm BB} = (w\fc^4)\ l \  F_{\rm Edd}. 
\ee

Therefore, the relation between the observed blackbody normalization $K$ and the NS radius, distance, and the spectral fitting parameters also changes:
\be
(w\fc^4)^{-1}\, T_{\rm BB}^4\, K =  T_\infty^4 \,\Omega,
\ee
or
\be \label{eq:KOmega}
     K = w\,\Omega.
\ee
Therefore, formally we have to fit the observed dependences $K^{-1/4} - F_{\rm BB}$ with the computed relations $w^{-1/4} - w\fc^4\,\ell$,
or simply just fitting the observed  $K - F_{\rm BB}$ dependency by $w - w\fc^4\,\ell$  (see equations \ref{eq:Fbb_l} and \ref{eq:KOmega} and 
Fig.~\ref{fig:dilut}). 
From here on, in order to simplify the presentation, we use the latter dependence.

There are two fitting parameters for this procedure,  $F_{\rm Edd}$ and $\Omega$. 
They can be combined to get the observed Eddington temperature $T_{\rm Edd,\infty}$ given by equation (\ref{eq:tedd}): 
\be \label{eq:teddN}
  T_{\rm Edd, \infty} = 9.81\, \left( F_{\rm Edd, -7}/\Omega \mbox{[(km/10~kpc)}^{2}] \right) ^{1/4}\,{\rm keV}. 
\ee
 The curves of equal $T_{\rm Edd,\infty}$ on the $M-R$ plane allow to limit  the NS mass and radius. 
 The second parameter, $\Omega$, could be useful if the distance is known,
otherwise the distance can be estimated using its value. 

\begin{figure} 
\begin{center} 
\includegraphics[width= 1.\columnwidth]{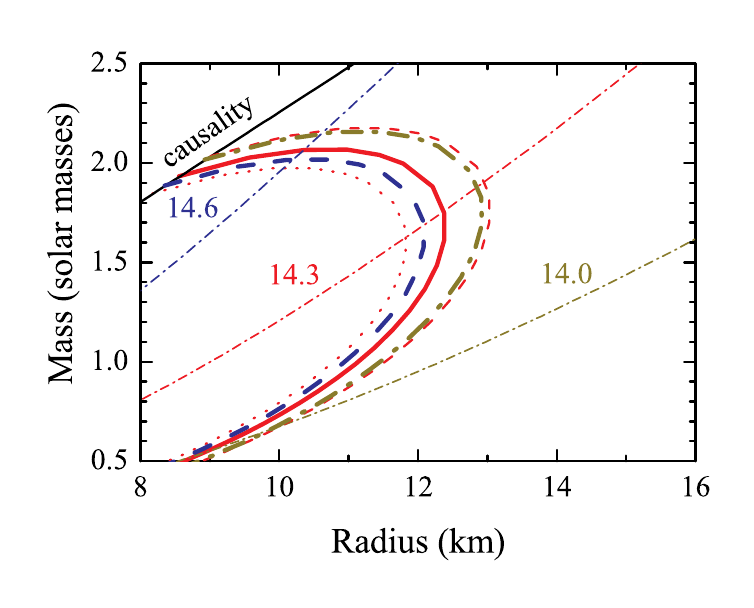}
\caption{\label{fig:tedd} 
Curves of constant $T_{\rm Edd,\infty}$ on the $M-R$ plane, obtained from the data with $F_{\min}=0.2F_{\rm td}$ 
and three different  $\log g=$14.6 (dashed blue curve), 14.3 (solid  red curve), and 14.0 (dot-dashed brown curve). 
The best fit parameters are given in Table~\ref{table:1}. 
The  red thin dotted curve corresponds to the best-fitting $T_{\rm Edd,\infty}$ from \citet{nattila16} for the same data.
The red thin dashed curve  corresponds to $F_{\min}=0.4F_{\rm td}$ and $\log g =14.3$. 
The constant $\log g$ curves are shown by the thin dot-dashed lines.  
} 
\end{center} 
\end{figure}

To illustrate the method  and demonstrate the differences between the new and the old approaches, we use the observational data of a  PRE burst from SAX\,J1810.8--2609 analysed by \citet{nattila16}. We leave the analysis of other sources for the future. 
To allow an easy comparison with previous results, we select the data points after the touchdown down to the minimum flux $F_{\min}$ corresponding to 20 per cent of the touchdown flux $F_{\rm td}$ (see Fig.~\ref{fig:ktbb}). To evaluate the effect of the data selection we also 
considered $F_{\min}=0.4F_{\rm td}$.   
The best-fitting solution was found with 
 the simple \citet{dem43} regression method by minimizing the function  
\be \label{eq:chi2}
       \chi^2 \!  \! =  \!  \! \sum_{i=1}^{N_{\rm obs}}\left[ \frac {(w\,\Omega - K_i)^2} {\Delta K_i^2}+ \frac{(w\fc^4\,\ell\,F_{\rm Edd,-7} - F_{\rm BB,i})^2}{\Delta F_{\rm BB,i}^2}\right] ,      
\ee
which  takes into account  errors on both axes. 
The term in the square brackets is the square of the minimal dimensionless distance from the given observed point to the model curve $w -w\fc^4\,\ell$ with the given fitting parameters $\Omega$ and $F_{\rm Edd,-7}$. 
The curve of constant $T_{\rm Edd,\infty}$ corresponding to the best-fitting parameters for $\log g=$14.3 is shown in  Fig.~\ref{fig:tedd} by the solid red curve, while similar curve obtained with the standard cooling tail method (see Table~1 in \citealt{nattila16}) is shown with the dotted red curve. 
It is clear that  the new method  gives NS radii larger by about 0.5~km.  
The reason is clear from equations~(\ref{eq:Fbb_l}) and (\ref{eq:KOmega}). 
The inverse bolometric correction $w\fc^4$ stretches the model curve, shifting it to higher fluxes at $\ell \gtrsim 0.9$ and 
to lower fluxes for $\ell\lesssim 0.9$ (see Fig.~\ref{fig:bolo}). In order to fit the data, we would need to decrease $F_{\rm Edd}$ and 
at the same time to increase $\Omega$. 
Both of these changes lead to smaller $T_{\rm Edd,\infty}$ and shift the corresponding solution to larger NS radii on the $M-R$ plane.

\begin{table}
\centering   
\caption{Parameters of the fits of the $K - F_{\rm BB}$ dependency with the corrected cooling tail models for various $\log g$. 
\label{table:1} 
}
\begin{tabular}[c]{c c c c  c }
\hline
$\log g$ & $F_{\rm Edd}$ 				        &  $\Omega$     & $T_{\rm Edd, \infty}$           & $\chi^2$/dof  \\ 
             &      $10^{-7}$ erg cm$^{-2}$ s$^{-1}$  & (km/10~kpc)$^{2}$  & (keV)  &    \\ 
\hline 
\multicolumn{5}{c}{$F_{\min}=0.2F_{\rm td}$ }  \\
14.6 &  0.771  & 1222 & 1.555 & 58.6/33 \\
14.3 &  0.776  & 1261 & 1.545 & 45.8/33 \\
14.0 &  0.776  & 1315 & 1.529 & 44.5/33 \\
 \hline
 \multicolumn{5}{c}{$F_{\min}=0.4F_{\rm td}$ }  \\
14.6 &  0.753  & 1293 & 1.524 & 30.6/17 \\ 
14.3 &  0.765  & 1308 & 1.525 & 28.8/17  \\
14.0 &  0.771  & 1338 & 1.520 & 27.8/17  \\
 \hline
\end{tabular}
\end{table}

The computed curves $\fc - \ell$ (as well as the curves $w -w\fc^4\,\ell$) are slightly different for  different $\log g$ \citep{SPW11,SPW12,Netal15}.  
Therefore, if we fit the observed curve $K - F_{\rm BB}$ with the theoretical dependences computed for different $\log g$,
we will obtain slightly different fitting parameters $\Omega$ and $F_{\rm Edd,-7}$. 
 As a result, the observed Eddington temperature $T_{\rm Edd,\infty}$ will be also different together with corresponding curves at the $M-R$ plane (see Fig.~\ref{fig:tedd}
 and Table~\ref{table:1}). 
In fact, the constant $T_{\rm Edd, \infty}$  curves corresponding to different $\log g$ 
give correct values of NS $M$ and $R$ only at the crossing points with the corresponding constant $\log g$ curves. 
Fortunately, all constant $T_{\rm Edd,\infty}$ curves are close to each other within  the statistical uncertainties.
 
We note here that the solution depends slightly on the number of  data
 points used in the fits and the method to find the best-fitting solution. 
For example,  if we take the points above $F_{\min}=0.4F_{\rm td}$, 
the constant $T_{\rm Edd,\infty}$ curve corresponding to the best-fitting  $\chi^2$ solution shifts to larger NS radii 
(dashed red curve in Fig.~\ref{fig:tedd} and Table~\ref{table:1}).

\section{Direct cooling tail method}
\label{sec:direct}

Formally, the cooling tail method gives a correct result only at the crossing point of the curves of the constant $\log g$ and the
corresponding $T_{\rm Edd, \infty}$.  
Therefore, we suggest here a direct cooling tail method with $M$ and $R$ as fitting parameters.
This method is free from the weakness of the standard method  \citep[see also][for more discussion]{Ozel15} 
and allows us to obtain solutions at the $R=2R_{\rm S}$ line as well. 
Moreover, the new formalism gives a possibility to take into account different priors in the physical parameters $M$ and $R$, and, for example, to control the minimum (or maximum) allowed mass in the fit.
 
For every pair $M$ and $R$  we can compute the gravity $\log g$, surface gravitational redshift $z$, and (assuming some  chemical composition) 
the observed Eddington temperature $T_{\rm Edd, \infty}$ (see equations (\ref{g}), (\ref{z}),  and (\ref{eq:teddN})). 
We use  interpolation in the sets of $\fc - \ell$ and $w\fc^4 -\ell$ curves, 
computed for different values of $\log g$, for finding the theoretical curve $w - w\fc^4\,\ell$  corresponding to a given $\log g$.

\subsection{Method of interpolation}

To improve the accuracy of the interpolation, we have extended the existing set of hot NS model atmospheres 
for $\log g$ = 14.0, 14.3, and 14.6 \citep{SPW12} to values from 13.7 up to 14.9 with the step 0.15 (i.e. nine values). 
We use the  same relative luminosity set $\ell$ as in the previous work of \citet{SPW12}, but for $\ell \ge 0.1$ only. 
Such computations were performed
for four chemical compositions, pure hydrogen, pure helium, and solar H/He mix with solar ($Z=Z_\odot$) and sub-solar ($Z=0.01Z_\odot$) metal  abundances. The model emergent spectra were fitted with a diluted blackbody (see details in \citealt{SPW11}) and corresponding
values of $\fc$, $w\fc^{4}$, and of the radiative acceleration $g_{\rm rad}$ were obtained.

Spectral fit parameters $\fc$ and $w$ depend on the relative luminosity $\ell$ in a similar way at different
$\log g$, when $\ell$ is small enough ($\ell < 0.8$). But  significant differences arise at larger $\ell$ because real electron scattering
opacity decreases  when the  plasma temperature increases (see \citealt{SPW12}). As a result, 
the maximum possible luminosity relative to the Eddington luminosity for Thomson opacity,  $\ell$, also increases.
Therefore, it is natural to use dependences of the type $\fc - g_{\rm rad}/g$ for interpolation, which are similar to each other for all surface gravities
(see Fig.~8 in \citealt{SPW12}).
At the first step, we find $g_{\rm rad}/g$ values corresponding to the $\ell$ grid. 
Then we introduce the fixed grid of $g_{\rm rad}/g$ 
and interpolate  all the necessary parameters ($T_{\rm eff}$, $\fc$, and $w$), to the new grid $g_{\rm rad}/g$ at every $\log g$. 
At the final step, we get the theoretical dependence  $w -  w\fc^4\,\ell$ for the $\log g$ computed for the investigated $M$ and $R$ pair
by interpolating all the quantities on the grid of nine values of  $\log g$.  
In this work we use the interpolation along weighting backward and forward parabolas \cite[see detail in][]{Kurucz70}.  

\subsection{Fitting procedure}

For the corrected cooling tail method we can use two fitting parameters, $\Omega$ and $F_{\rm Edd,-7}$. 
However, for a given $M$ and $R$ pair,  $L_{\rm Edd}$ is already determined by their values, and  both 
fitting parameters depend on only one parameter -- the distance to the source $D$.
 This allows us to obtain an estimation of the distance to the source given the chemical composition of the 
NS atmosphere and to limit the final solution if there exist independent  constraints on the distance.  

In the direct cooling tail method, we minimize $\chi^2$ at a grid in the $M-R$ plane 
by fitting the observed $K - F_{\rm BB}$ curve 
with the  theoretical $w-w\fc^4\,\ell$ curve interpolated to the current $\log g$
using  $D_{10}$ as the only fitting parameter. 
We consider a grid of masses from 1 to 3$\Msun$ with the step 0.01$\Msun$ and  
a grid of radii from 9  to 17 km with the step 0.01 km. 
We ignore the pairs which do not satisfy the causality condition $R>1.45R_{\rm S}$ \citep{LP07}.
We thus obtain the map of best-fitting $D$ and the $\chi^2$ values at the $M-R$ plane and 
the overall minimum of  $\chi^2$, which allow us to find the confidence regions 
(e.g. for two parameters, $\Delta\chi^2=2.3, 4.61, 9.21$ give probabilities of 68, 90 and 99\%).
We note here that, the method used for finding the best-fitting solution, e.g. the $\chi^2$-method like here and in \citet{SPRW11}, or the robust likelihood method from 
\citet{Poutanen.etal:14},  affects only slightly (within the 1$\sigma$ error) the centroid of the solution. 
On the other hand, the resulting error contours on the $M-R$ plane are affected more: the $\chi^2$-method generally underestimates the errors, 
 because of the presence of the significant scattering in the data 
 as well as because the best-fitting solution may be located at the boundary of the prior distributions.
The robust estimator gives more correct presentation of the errors when benchmarked against a full Bayesian fit with intrinsic scatter in the system.    
For simplicity we, however, here show only the results obtained by minimizing a non-robust $\chi^2$ function given by equation (\ref{eq:chi2}).

\begin{figure} 
\begin{center}  
\includegraphics[width= 1.\columnwidth]{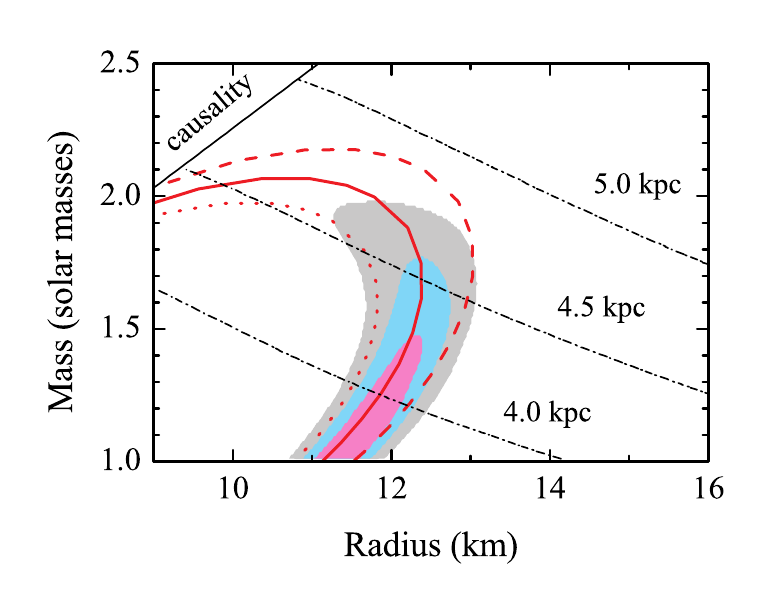}
\caption{\label{fig:mr_short} 
The $\chi^2$ confidence regions (68, 90 and 99\% probabilities) in the mass-radius plane for
SAX\,J1810.8--2609 obtained using the  direct cooling tail method.  
The data above $F_{\min}=0.2F_{\rm td}$  (see Fig.~\ref{fig:ktbb}) are used. 
The red solid and dashed curves  correspond to the best-fitting $T_{\rm Edd, \infty}$ obtained for $\log g$=14.3 
for $F_{\min}=0.2F_{\rm td}$  and $0.4F_{\rm td}$, respectively, 
while the red dotted curve gives the results from \citet{nattila16}. 
The black dot-dashed curves correspond to the constant distance of 4.0, 4.5, and 5.0 kpc.
} 
\end{center} 
\end{figure}

We apply the direct cooling tail method to the data described above and presented in Fig.~\ref{fig:ktbb}. 
The minimum  $\chi^2$  is 45.4 for 32 dof.  
The confidence regions at the $M-R$ plane are shown in Fig.~\ref{fig:mr_short}. 
To compare with the results presented in Fig.~\ref{fig:tedd} we also draw the curve of equal $T_{\rm Edd, \infty}$ obtained 
using the corrected cooling tail method for $\log g$= 14.3 (red solid curve).  
There is good correspondence between this curve and the confidence regions given by the direct method. 
To compare with the previous result, the curve of equal $T_{\rm Edd, \infty}$ obtained with the standard cooling tail method 
\citep[Table 2 in][]{nattila16} is also shown (red dotted  curve).  
We remind here that the method does not suffer from the problems with the Jacobian transformation  from $F_{\rm Edd},\Omega$ to 
$M,R$ \citep[see discussion in ][]{Ozel15}, because we fit directly on the $M-R$ plane. 
We also see that instead of the two regions of allowed solutions (small $R$, large $M$ and large $R$, smaller $M$; see e.g. \citealt{SPRW11,Poutanen.etal:14}) the  solution here prefers larger $R$ and smaller $M$ values. 
This is a direct consequence of the fact that the shape of the  $w-w\fc^4\,\ell$ curves depends on $\log g$ and a smaller $\log g$ gives a better description of the data
(see Table~\ref{table:1}).

We conclude that the most probable radius of the NS in SAX\,J1810.8--2609 lies in the range between 11.5  and 13.0~km for the mass range of 1.3--1.8$\Msun$.
However, the smaller radii of $\sim$10~km for a high-mass  ($\sim 2 \Msun$)  NS cannot be formally excluded.
The obtained constraints also limit the possible distance to the source, which likely lies in the interval between 3.7 and 4.7~kpc. 

\section{Summary}
\label{sec:summary}
 
We have presented the direct cooling tail method, which takes NS mass and radius together with the distance as the direct fitting parameters.
This method implies fitting of an observed dependence between  blackbody fitting parameters, the normalization and the 
observed blackbody flux, $K - F_{\rm BB}$, with the corresponding theoretical dependence, $w - w\fc^4\,\ell$, 
which is found for every given $M$ and $R$ pair by interpolation in the extended grid (for nine values of $\log g$) of the computed dependences $\fc - \ell$ and $w\fc^4 - \ell$.
Such grids were computed for 4 chemical compositions: pure H and pure He, and solar H/He mix with the solar ($Z=Z_\odot$) and sub-solar ($Z=0.01Z_\odot$) metal abundances. 
The fitting procedure results in the $\chi^2$ maps and corresponding confidence regions in the $M-R$ plane. 
Other fitting methods (robust likelihood, Bayesian) give very similar results with slightly larger errors. 
Compared with the standard cooling tail method, the new method does not suffer from the problems with the Jacobian transformation from $(F_{\rm Edd},\Omega)$ to 
$(M,R)$ allowing to obtain solution anywhere on the $M-R$ plane.

We applied the new method to a PRE burst that occurred in the low/hard state of SAX J1810.8--2609.  
The direct cooling tail method gives the Eddington temperature slightly smaller than the standard cooling tail method, 
resulting in a  systematic shift of the radii by $\sim 1\,\sigma$  to larger values (which is still within the statistical uncertainties). 
However, instead of the two separated $(M,R)$ regions usually obtained with the standard method, our solution prefers larger $R$ and smaller $M$ values corresponding to 
smaller $\log g$ values, which is a direct consequence of the dependence of the theoretical cooling curves $w - w\fc^4\,\ell$ on gravity. 
We finally constrain the NS radius in  SAX\,J1810.8--2609 to lie between 11.5  and 13.0 km for the assumed mass range of 1.3--1.8$\Msun$. 
The distance to the source likely lies between 3.7 and 4.7~kpc.

\section*{Acknowledgments} 
The work was supported by the German Research Foundation (DFG) grant WE 1312/48-1, 
the Magnus Ehrnrooth Foundation,  the Russian Foundation for Basic Research  grant {16-02-01145-a} (VFS), 
the Foundations' Professor Pool (JP), the Finnish Cultural Foundation (JP), the Academy of Finland grant 268740 (JP, JJEK), 
the University of Turku Graduate School in Physical and Chemical Sciences (JN), 
the Faculty of the European Space Astronomy Centre (JN, JJEK), the European Space Agency research fellowship programm
(JJEK), and the Russian Science Foundation grant 14-12-01287 (MGR). 
We also acknowledge the support from the International Space Science Institute (Bern, Switzerland) 
and  COST Action MP1304 NewCompStar.       



\bibliographystyle{mnras}
\bibliography{xrb} % if your bibtex file is called example.bib

\begin{thebibliography}{}
\makeatletter
\relax
\def\mn@urlcharsother{\let\do\@makeother \do\$\do\&\do\#\do\^\do\_\do\%\do\~}
\def\mn@doi{\begingroup\mn@urlcharsother \@ifnextchar [ {\mn@doi@}
  {\mn@doi@[]}}
\def\mn@doi@[#1]#2{\def\@tempa{#1}\ifx\@tempa\@empty \href
  {http://dx.doi.org/#2} {doi:#2}\else \href {http://dx.doi.org/#2} {#1}\fi
  \endgroup}
\def\mn@eprint#1#2{\mn@eprint@#1:#2::\@nil}
\def\mn@eprint@arXiv#1{\href {http://arxiv.org/abs/#1} {{\tt arXiv:#1}}}
\def\mn@eprint@dblp#1{\href {http://dblp.uni-trier.de/rec/bibtex/#1.xml}
  {dblp:#1}}
\def\mn@eprint@#1:#2:#3:#4\@nil{\def\@tempa {#1}\def\@tempb {#2}\def\@tempc
  {#3}\ifx \@tempc \@empty \let \@tempc \@tempb \let \@tempb \@tempa \fi \ifx
  \@tempb \@empty \def\@tempb {arXiv}\fi \@ifundefined
  {mn@eprint@\@tempb}{\@tempb:\@tempc}{\expandafter \expandafter \csname
  mn@eprint@\@tempb\endcsname \expandafter{\@tempc}}}

\bibitem[\protect\citeauthoryear{{Antoniadis} et~al.,}{{Antoniadis}
  et~al.}{2013}]{Anton.etal:13}
{Antoniadis} J.,  et~al., 2013, \mn@doi [Science] {10.1126/science.1233232},
  \href {http://adsabs.harvard.edu/abs/2013Sci...340..448A} {340, 448}

\bibitem[\protect\citeauthoryear{{Damen}, {Magnier}, {Lewin}, {Tan}, {Penninx}
  \& {van Paradijs}}{{Damen} et~al.}{1990}]{Damen:90}
{Damen} E.,  {Magnier} E.,  {Lewin} W.~H.~G.,  {Tan} J.,  {Penninx} W.,   {van
  Paradijs} J.,  1990, \aap, \href
  {http://ads.ari.uni-heidelberg.de/abs/1990A\&A...237..103D} {237, 103}

\bibitem[\protect\citeauthoryear{{Deming}}{{Deming}}{2011}]{dem43}
{Deming} W.~E.,  2011, {Statistical Adjustment of Data (Dover Books on
  Mathematics)}.
Dover Publications, New York

\bibitem[\protect\citeauthoryear{{Ebisuzaki}}{{Ebisuzaki}}{1987}]{Ebisuzaki87}
{Ebisuzaki} T.,  1987, \pasj, \href
  {http://adsabs.harvard.edu/cgi-bin/nph-bib_query?bibcode=1998A%26A...340..61%
7M&db_key=AST} {39, 287}

\bibitem[\protect\citeauthoryear{{Galloway}, {Muno}, {Hartman}, {Psaltis}  \&
  {Chakrabarty}}{{Galloway} et~al.}{2008}]{gallow08}
{Galloway} D.~K.,  {Muno} M.~P.,  {Hartman} J.~M.,  {Psaltis} D.,
  {Chakrabarty} D.,  2008, \mn@doi [\apjs] {10.1086/592044}, \href
  {http://adsabs.harvard.edu/abs/2008ApJS..179..360G} {179, 360}

\bibitem[\protect\citeauthoryear{{Guillot}, {Servillat}, {Webb}  \&
  {Rutledge}}{{Guillot} et~al.}{2013}]{Gulliot.etal:13}
{Guillot} S.,  {Servillat} M.,  {Webb} N.~A.,   {Rutledge} R.~E.,  2013,
  \mn@doi [\apj] {10.1088/0004-637X/772/1/7}, \href
  {http://adsabs.harvard.edu/abs/2013ApJ...772....7G} {772, 7}

\bibitem[\protect\citeauthoryear{{G{\"u}ver}, {Psaltis}  \&
  {{\"O}zel}}{{G{\"u}ver} et~al.}{2012}]{Guver.etal:12}
{G{\"u}ver} T.,  {Psaltis} D.,   {{\"O}zel} F.,  2012, \mn@doi [\apj]
  {10.1088/0004-637X/747/1/76}, \href
  {http://adsabs.harvard.edu/abs/2012ApJ...747...76G} {747, 76}

\bibitem[\protect\citeauthoryear{{Haensel}, {Potekhin}  \&
  {Yakovlev}}{{Haensel} et~al.}{2007}]{HPY:07}
{Haensel} P.,  {Potekhin} A.~Y.,   {Yakovlev} D.~G.,  2007, {Neutron Stars 1:
  Equation of State and Structure}.
 Astrophysics and Space Science Library Vol. 326, Springer, New York

\bibitem[\protect\citeauthoryear{{Heinke}, {Rybicki}, {Narayan}  \&
  {Grindlay}}{{Heinke} et~al.}{2006}]{HR06}
{Heinke} C.~O.,  {Rybicki} G.~B.,  {Narayan} R.,   {Grindlay} J.~E.,  2006,
  \mn@doi [\apj] {10.1086/503701}, \href
  {http://adsabs.harvard.edu/abs/2006ApJ...644.1090H} {644, 1090}

\bibitem[\protect\citeauthoryear{{Heinke} et~al.,}{{Heinke}
  et~al.}{2014}]{Heinke.etal:14}
{Heinke} C.~O.,  et~al., 2014, \mn@doi [\mnras] {10.1093/mnras/stu1449}, \href
  {http://adsabs.harvard.edu/abs/2014MNRAS.444..443H} {444, 443}

\bibitem[\protect\citeauthoryear{{Ho} \& {Heinke}}{{Ho} \&
  {Heinke}}{2009}]{HH:09}
{Ho} W.~C.~G.,  {Heinke} C.~O.,  2009, \mn@doi [\nat] {10.1038/nature08525},
  \href {http://adsabs.harvard.edu/abs/2009Natur.462...71H} {462, 71}

\bibitem[\protect\citeauthoryear{{Kajava} et~al.,}{{Kajava}
  et~al.}{2014}]{Kajava.etal:14}
{Kajava} J.~J.~E.,  et~al., 2014, \mn@doi [\mnras] {10.1093/mnras/stu2073},
  \href {http://adsabs.harvard.edu/abs/2014MNRAS.445.4218K} {445, 4218}

\bibitem[\protect\citeauthoryear{{Kajava}, {N{\"a}ttil{\"a}}, {Poutanen},
  {Cumming}, {Suleimanov}  \& {Kuulkers}}{{Kajava} et~al.}{2017}]{Kajava17}
{Kajava} J.~J.~E.,  {N{\"a}ttil{\"a}} J.,  {Poutanen} J.,  {Cumming} A.,
  {Suleimanov} V.,   {Kuulkers} E.,  2017, \mn@doi [\mnras]
  {10.1093/mnrasl/slw167}, \href
  {http://adsabs.harvard.edu/abs/2017MNRAS.464L...6K} {464, L6}

\bibitem[\protect\citeauthoryear{{Kiziltan}, {Kottas}, {De Yoreo}  \&
  {Thorsett}}{{Kiziltan} et~al.}{2013}]{Kiziltan13}
{Kiziltan} B.,  {Kottas} A.,  {De Yoreo} M.,   {Thorsett} S.~E.,  2013, \mn@doi
  [\apj] {10.1088/0004-637X/778/1/66}, \href
  {http://adsabs.harvard.edu/abs/2013ApJ...778...66K} {778, 66}

\bibitem[\protect\citeauthoryear{{Klochkov}, {P{\"u}hlhofer}, {Suleimanov},
  {Simon}, {Werner}  \& {Santangelo}}{{Klochkov}
  et~al.}{2013}]{Klochkov.etal:13}
{Klochkov} D.,  {P{\"u}hlhofer} G.,  {Suleimanov} V.,  {Simon} S.,  {Werner}
  K.,   {Santangelo} A.,  2013, \mn@doi [\aap] {10.1051/0004-6361/201321740},
  \href {http://adsabs.harvard.edu/abs/2013A%26A...556A..41K} {556, A41}

\bibitem[\protect\citeauthoryear{{Klochkov}, {Suleimanov}, {P{\"u}hlhofer},
  {Yakovlev}, {Santangelo}  \& {Werner}}{{Klochkov}
  et~al.}{2015}]{Klochkov.etal:15}
{Klochkov} D.,  {Suleimanov} V.,  {P{\"u}hlhofer} G.,  {Yakovlev} D.~G.,
  {Santangelo} A.,   {Werner} K.,  2015, \mn@doi [\aap]
  {10.1051/0004-6361/201424683}, \href
  {http://adsabs.harvard.edu/abs/2015A%26A...573A..53K} {573, A53}

\bibitem[\protect\citeauthoryear{{Koljonen}, {Kajava}  \&
  {Kuulkers}}{{Koljonen} et~al.}{2016}]{Koljonen16}
{Koljonen} K.~I.~I.,  {Kajava} J.~J.~E.,   {Kuulkers} E.,  2016, \mn@doi [\apj]
  {10.3847/0004-637X/829/2/91}, \href
  {http://adsabs.harvard.edu/abs/2016ApJ...829...91K} {829, 91}

\bibitem[\protect\citeauthoryear{{Kramer} \& {Wex}}{{Kramer} \&
  {Wex}}{2009}]{Kramer09}
{Kramer} M.,  {Wex} N.,  2009, \mn@doi [Classical and Quantum Gravity]
  {10.1088/0264-9381/26/7/073001}, \href
  {http://adsabs.harvard.edu/abs/2009CQGra..26g3001K} {26, 073001}

\bibitem[\protect\citeauthoryear{{Kurucz}}{{Kurucz}}{1970}]{Kurucz70}
{Kurucz} R.~L.,  1970, SAO Special Report, \href
  {http://adsabs.harvard.edu/abs/1970SAOSR.309.....K} {309}

\bibitem[\protect\citeauthoryear{{Ku{\'s}mierek}, {Madej}  \&
  {Kuulkers}}{{Ku{\'s}mierek} et~al.}{2011}]{Kusm:11}
{Ku{\'s}mierek} K.,  {Madej} J.,   {Kuulkers} E.,  2011, \mn@doi [\mnras]
  {10.1111/j.1365-2966.2011.18948.x}, \href
  {http://ads.ari.uni-heidelberg.de/abs/2011MNRAS.415.3344K} {415, 3344}

\bibitem[\protect\citeauthoryear{{Lattimer} \& {Prakash}}{{Lattimer} \&
  {Prakash}}{2007}]{LP07}
{Lattimer} J.~M.,  {Prakash} M.,  2007, \mn@doi [\physrep]
  {10.1016/j.physrep.2007.02.003}, \href
  {http://adsabs.harvard.edu/abs/2007PhR...442..109L} {442, 109}

\bibitem[\protect\citeauthoryear{{Lattimer} \& {Steiner}}{{Lattimer} \&
  {Steiner}}{2014}]{LS:14}
{Lattimer} J.~M.,  {Steiner} A.~W.,  2014, \mn@doi [\apj]
  {10.1088/0004-637X/784/2/123}, \href
  {http://adsabs.harvard.edu/abs/2014ApJ...784..123L} {784, 123}

\bibitem[\protect\citeauthoryear{{Lewin}, {van Paradijs}  \& {Taam}}{{Lewin}
  et~al.}{1993}]{LvPT93}
{Lewin} W.~H.~G.,  {van Paradijs} J.,   {Taam} R.~E.,  1993, \mn@doi [Space
  Science Reviews] {10.1007/BF00196124}, \href
  {http://adsabs.harvard.edu/abs/1993SSRv...62..223L} {62, 223}

\bibitem[\protect\citeauthoryear{{Lo}, {Miller}, {Bhattacharyya}  \&
  {Lamb}}{{Lo} et~al.}{2013}]{Lo13}
{Lo} K.~H.,  {Miller} M.~C.,  {Bhattacharyya} S.,   {Lamb} F.~K.,  2013,
  \mn@doi [\apj] {10.1088/0004-637X/776/1/19}, \href
  {http://adsabs.harvard.edu/abs/2013ApJ...776...19L} {776, 19}

\bibitem[\protect\citeauthoryear{{Miller}}{{Miller}}{2013}]{Miller13}
{Miller} M.~C.,  2013, preprint, \href
  {http://adsabs.harvard.edu/abs/2013arXiv1312.0029M} {} (\mn@eprint {arXiv}
  {1312.0029})

\bibitem[\protect\citeauthoryear{{Miller} \& {Lamb}}{{Miller} \&
  {Lamb}}{2015}]{ML15}
{Miller} M.~C.,  {Lamb} F.~K.,  2015, \mn@doi [\apj]
  {10.1088/0004-637X/808/1/31}, \href
  {http://adsabs.harvard.edu/abs/2015ApJ...808...31M} {808, 31}

\bibitem[\protect\citeauthoryear{{Miller} \& {Lamb}}{{Miller} \&
  {Lamb}}{2016}]{ML16}
{Miller} M.~C.,  {Lamb} F.~K.,  2016, \mn@doi [European Physical Journal A]
  {10.1140/epja/i2016-16063-8}, \href
  {http://adsabs.harvard.edu/abs/2016EPJA...52...63M} {52, 63}

\bibitem[\protect\citeauthoryear{{N{\"a}ttil{\"a}}, {Suleimanov}, {Kajava}  \&
  {Poutanen}}{{N{\"a}ttil{\"a}} et~al.}{2015}]{Netal15}
{N{\"a}ttil{\"a}} J.,  {Suleimanov} V.~F.,  {Kajava} J.~J.~E.,   {Poutanen} J.,
   2015, \mn@doi [\aap] {10.1051/0004-6361/201526512}, \href
  {http://adsabs.harvard.edu/abs/2015A%26A...581A..83N} {581, A83}

\bibitem[\protect\citeauthoryear{{N{\"a}ttil{\"a}}, {Steiner}, {Kajava},
  {Suleimanov}  \& {Poutanen}}{{N{\"a}ttil{\"a}} et~al.}{2016}]{nattila16}
{N{\"a}ttil{\"a}} J.,  {Steiner} A.~W.,  {Kajava} J.~J.~E.,  {Suleimanov}
  V.~F.,   {Poutanen} J.,  2016, \mn@doi [\aap]
  {http://dx.doi.org/10.1051/0004-6361/201527416}, \href
  {http://adsabs.harvard.edu/abs/2015arXiv150906561N} {591, A25}

\bibitem[\protect\citeauthoryear{{{\"O}zel}}{{{\"O}zel}}{2006}]{Ozel06}
{{\"O}zel} F.,  2006, \mn@doi [\nat] {10.1038/nature04858}, \href
  {http://adsabs.harvard.edu/abs/2006Natur.441.1115O} {441, 1115}

\bibitem[\protect\citeauthoryear{{{\"O}zel}}{{{\"O}zel}}{2013}]{Ozel13}
{{\"O}zel} F.,  2013, \mn@doi [Reports on Progress in Physics]
  {10.1088/0034-4885/76/1/016901}, \href
  {http://adsabs.harvard.edu/abs/2013RPPh...76a6901O} {76, 016901}

\bibitem[\protect\citeauthoryear{{{\"O}zel} \& {Psaltis}}{{{\"O}zel} \&
  {Psaltis}}{2015}]{Ozel15}
{{\"O}zel} F.,  {Psaltis} D.,  2015, \mn@doi [\apj]
  {10.1088/0004-637X/810/2/135}, \href
  {http://adsabs.harvard.edu/abs/2015ApJ...810..135O} {810, 135}

\bibitem[\protect\citeauthoryear{{{\"O}zel}, {G{\"u}ver}  \&
  {Psaltis}}{{{\"O}zel} et~al.}{2009}]{ozel:09}
{{\"O}zel} F.,  {G{\"u}ver} T.,   {Psaltis} D.,  2009, \mn@doi [\apj]
  {10.1088/0004-637X/693/2/1775}, \href
  {http://adsabs.harvard.edu/abs/2009ApJ...693.1775O} {693, 1775}

\bibitem[\protect\citeauthoryear{{{\"O}zel}, {Psaltis}, {Narayan}  \& {Santos
  Villarreal}}{{{\"O}zel} et~al.}{2012}]{OPN12}
{{\"O}zel} F.,  {Psaltis} D.,  {Narayan} R.,   {Santos Villarreal} A.,  2012,
  \mn@doi [\apj] {10.1088/0004-637X/757/1/55}, \href
  {http://adsabs.harvard.edu/abs/2012ApJ...757...55O} {757, 55}

\bibitem[\protect\citeauthoryear{{{\"O}zel}, {Psaltis}, {G{\"u}ver}, {Baym},
  {Heinke}  \& {Guillot}}{{{\"O}zel} et~al.}{2016}]{Ozel16}
{{\"O}zel} F.,  {Psaltis} D.,  {G{\"u}ver} T.,  {Baym} G.,  {Heinke} C.,
  {Guillot} S.,  2016, \mn@doi [\apj] {10.3847/0004-637X/820/1/28}, \href
  {http://adsabs.harvard.edu/abs/2016ApJ...820...28O} {820, 28}

\bibitem[\protect\citeauthoryear{{Pavlov} \& {Luna}}{{Pavlov} \&
  {Luna}}{2009}]{PL:09}
{Pavlov} G.~G.,  {Luna} G.~J.~M.,  2009, \mn@doi [\apj]
  {10.1088/0004-637X/703/1/910}, \href
  {http://adsabs.harvard.edu/abs/2009ApJ...703..910P} {703, 910}

\bibitem[\protect\citeauthoryear{{Poutanen} \& {Gierli{\'n}ski}}{{Poutanen} \&
  {Gierli{\'n}ski}}{2003}]{PG03}
{Poutanen} J.,  {Gierli{\'n}ski} M.,  2003, \mn@doi [\mnras]
  {10.1046/j.1365-8711.2003.06773.x}, \href
  {http://adsabs.harvard.edu/abs/2003MNRAS.343.1301P} {343, 1301}

\bibitem[\protect\citeauthoryear{{Poutanen}, {N{\"a}ttil{\"a}}, {Kajava},
  {Latvala}, {Galloway}, {Kuulkers}  \& {Suleimanov}}{{Poutanen}
  et~al.}{2014}]{Poutanen.etal:14}
{Poutanen} J.,  {N{\"a}ttil{\"a}} J.,  {Kajava} J.~J.~E.,  {Latvala} O.-M.,
  {Galloway} D.~K.,  {Kuulkers} E.,   {Suleimanov} V.~F.,  2014, \mn@doi
  [\mnras] {10.1093/mnras/stu1139}, \href
  {http://adsabs.harvard.edu/abs/2014MNRAS.442.3777P} {442, 3777}

\bibitem[\protect\citeauthoryear{{Steiner}, {Lattimer}  \& {Brown}}{{Steiner}
  et~al.}{2010}]{SLB10}
{Steiner} A.~W.,  {Lattimer} J.~M.,   {Brown} E.~F.,  2010, \mn@doi [\apj]
  {10.1088/0004-637X/722/1/33}, \href
  {http://adsabs.harvard.edu/abs/2010ApJ...722...33S} {722, 33}

\bibitem[\protect\citeauthoryear{{Suleimanov} \& {Werner}}{{Suleimanov} \&
  {Werner}}{2007}]{sw:07}
{Suleimanov} V.,  {Werner} K.,  2007, \mn@doi [\aap]
  {10.1051/0004-6361:20066174}, \href
  {http://adsabs.harvard.edu/abs/2007A%26A...466..661S} {466, 661}

\bibitem[\protect\citeauthoryear{{Suleimanov}, {Poutanen}  \&
  {Werner}}{{Suleimanov} et~al.}{2011a}]{SPW11}
{Suleimanov} V.,  {Poutanen} J.,   {Werner} K.,  2011a, \mn@doi [\aap]
  {10.1051/0004-6361/201015845}, \href
  {http://adsabs.harvard.edu/abs/2011A%26A...527A.139S} {527, A139}

\bibitem[\protect\citeauthoryear{{Suleimanov}, {Poutanen}, {Revnivtsev}  \&
  {Werner}}{{Suleimanov} et~al.}{2011b}]{SPRW11}
{Suleimanov} V.,  {Poutanen} J.,  {Revnivtsev} M.,   {Werner} K.,  2011b,
  \mn@doi [\apj] {10.1088/0004-637X/742/2/122}, \href
  {http://adsabs.harvard.edu/abs/2011ApJ...742..122S} {742, 122}

\bibitem[\protect\citeauthoryear{{Suleimanov}, {Poutanen}  \&
  {Werner}}{{Suleimanov} et~al.}{2012}]{SPW12}
{Suleimanov} V.,  {Poutanen} J.,   {Werner} K.,  2012, \mn@doi [\aap]
  {10.1051/0004-6361/201219480}, \href
  {http://adsabs.harvard.edu/abs/2012A%26A...545A.120S} {545, A120}

\bibitem[\protect\citeauthoryear{{Suleimanov}, {Poutanen}, {Klochkov}  \&
  {Werner}}{{Suleimanov} et~al.}{2016}]{Suleimanov16EPJA}
{Suleimanov} V.~F.,  {Poutanen} J.,  {Klochkov} D.,   {Werner} K.,  2016,
  \mn@doi [European Physical Journal A] {10.1140/epja/i2016-16020-7}, \href
  {http://adsabs.harvard.edu/abs/2016EPJA...52...20S} {52, 20}

\bibitem[\protect\citeauthoryear{{Thorsett} \& {Chakrabarty}}{{Thorsett} \&
  {Chakrabarty}}{1999}]{TC99}
{Thorsett} S.~E.,  {Chakrabarty} D.,  1999, \mn@doi [\apj] {10.1086/306742},
  \href {http://adsabs.harvard.edu/abs/1999ApJ...512..288T} {512, 288}

\bibitem[\protect\citeauthoryear{{Watts} et~al.,}{{Watts}
  et~al.}{2016}]{watts16}
{Watts} A.~L.,  et~al., 2016, \mn@doi [Reviews of Modern Physics]
  {10.1103/RevModPhys.88.021001}, \href
  {http://adsabs.harvard.edu/abs/2016RvMP...88b1001W} {88, 021001}

\bibitem[\protect\citeauthoryear{{Worpel}, {Galloway}  \& {Price}}{{Worpel}
  et~al.}{2015}]{Worpel.etal:15}
{Worpel} H.,  {Galloway} D.~K.,   {Price} D.~J.,  2015, \mn@doi [\apj]
  {10.1088/0004-637X/801/1/60}, \href
  {http://adsabs.harvard.edu/abs/2015ApJ...801...60W} {801, 60}

\bibitem[\protect\citeauthoryear{{Zavlin}, {Pavlov}  \& {Shibanov}}{{Zavlin}
  et~al.}{1996}]{Zavlin.etal:96}
{Zavlin} V.~E.,  {Pavlov} G.~G.,   {Shibanov} Y.~A.,  1996, \aap, \href
  {http://adsabs.harvard.edu/abs/1996A%26A...315..141Z} {315, 141}

\bibitem[\protect\citeauthoryear{{Zavlin}, {Pavlov}  \& {Trumper}}{{Zavlin}
  et~al.}{1998}]{Zavlin.etal:98}
{Zavlin} V.~E.,  {Pavlov} G.~G.,   {Trumper} J.,  1998, \aap, \href
  {http://adsabs.harvard.edu/abs/1998A%26A...331..821Z} {331, 821}

\bibitem[\protect\citeauthoryear{{van Paradijs}, {Dotani}, {Tanaka}  \&
  {Tsuru}}{{van Paradijs} et~al.}{1990}]{vP:90}
{van Paradijs} J.,  {Dotani} T.,  {Tanaka} Y.,   {Tsuru} T.,  1990, \pasj,
  \href {http://ads.ari.uni-heidelberg.de/abs/1990PASJ...42..633V} {42, 633}

\makeatother
\end{thebibliography}

% Don't change these lines
\bsp	% typesetting comment
\label{lastpage}
\end{document}